\documentclass[12pt]{article}

\usepackage{graphicx}   
\usepackage{xcolor}

\usepackage{geometry}
 \usepackage{amsmath}

\makeatletter
\newsavebox{\@brx}
\newcommand{\llangle}[1][]{\savebox{\@brx}{\(\m@th{#1\langle}\)}%
  \mathopen{\copy\@brx\kern-0.5\wd\@brx\usebox{\@brx}}}
\newcommand{\rrangle}[1][]{\savebox{\@brx}{\(\m@th{#1\rangle}\)}%
  \mathclose{\copy\@brx\kern-0.5\wd\@brx\usebox{\@brx}}}
\makeatother

\newcommand{\rmd}{{\rm d}}

\usepackage{etoolbox}
\makeatletter
\newrobustcmd{\fixappendix}{%
  \patchcmd{\l@section}{1.5em}{7em}{}{}%
  \patchcmd{\l@subsection}{2.3em}{7em}{}{}%
}

\usepackage{sectsty}
\sectionfont{\fontsize{12}{15}\selectfont}
\subsectionfont{\fontsize{12}{15}\selectfont}

\begin{document}





\begin{minipage}{14cm}
\begin{center}
{\fontsize{13}{10}\selectfont {\textbf{Dynamics of inertial particles under velocity resetting}}}\\
\vspace{10pt}
{\fontsize{12}{10}\selectfont {Kristian St\o{}levik Olsen$^*$  and Hartmut L\"{o}wen}}\\
\vspace{4pt}
{\fontsize{10}{10}\selectfont \noindent \textit{ Institut für Theoretische Physik II - Weiche Materie, Heinrich-Heine-Universität Düsseldorf, D-40225 Düsseldorf, Germany}}\\
\vspace{4pt}
{\fontsize{9}{10}\selectfont \noindent ${\:}^*$\textit{e-mail: olsen@thphy.uni-duesseldorf.de}}
\end{center}
\end{minipage}

\vspace{5pt}

\begin{abstract}\fontsize{9}{10}\selectfont
\noindent\textbf{Abstract: }We investigate stochastic resetting in coupled systems involving two degrees of freedom, where only one variable is reset. The resetting variable, which we think of as hidden, indirectly affects the remaining observable variable through correlations. We derive the Fourier-Laplace transform of the observable variable's propagator and provide a recursive relation for all the moments, facilitating a comprehensive examination of the process. We apply this framework to inertial transport processes where we observe {\color{black} the} particle position while the velocity is hidden and is being reset at a constant rate. We show that velocity resetting results in a linearly growing spatial mean squared displacement at late times, independently of reset-free dynamics, due to resetting-induced tempering of velocity correlations. General expressions for the effective diffusion and drift coefficients are derived as function of resetting rate. Non-trivial dependence on the rate may appear due to multiple timescales and crossovers in the reset-free dynamics. An extension that incorporates refractory periods after each reset is considered, where the post-resetting pauses can lead to anomalous diffusive behavior. Our results are of relevance to a wide range of systems, including inertial transport where mechanical momentum is lost in collisions with the environment, or the behavior of living organisms where stop-and-go locomotion with inertia is ubiquitous. Numerical simulations for underdamped Brownian motion and the random acceleration process confirm our findings.
\end{abstract}

%
%
%
%
%

\section{Introduction}
\begin{figure}
    \centering
    \includegraphics[width = 0.7\columnwidth]{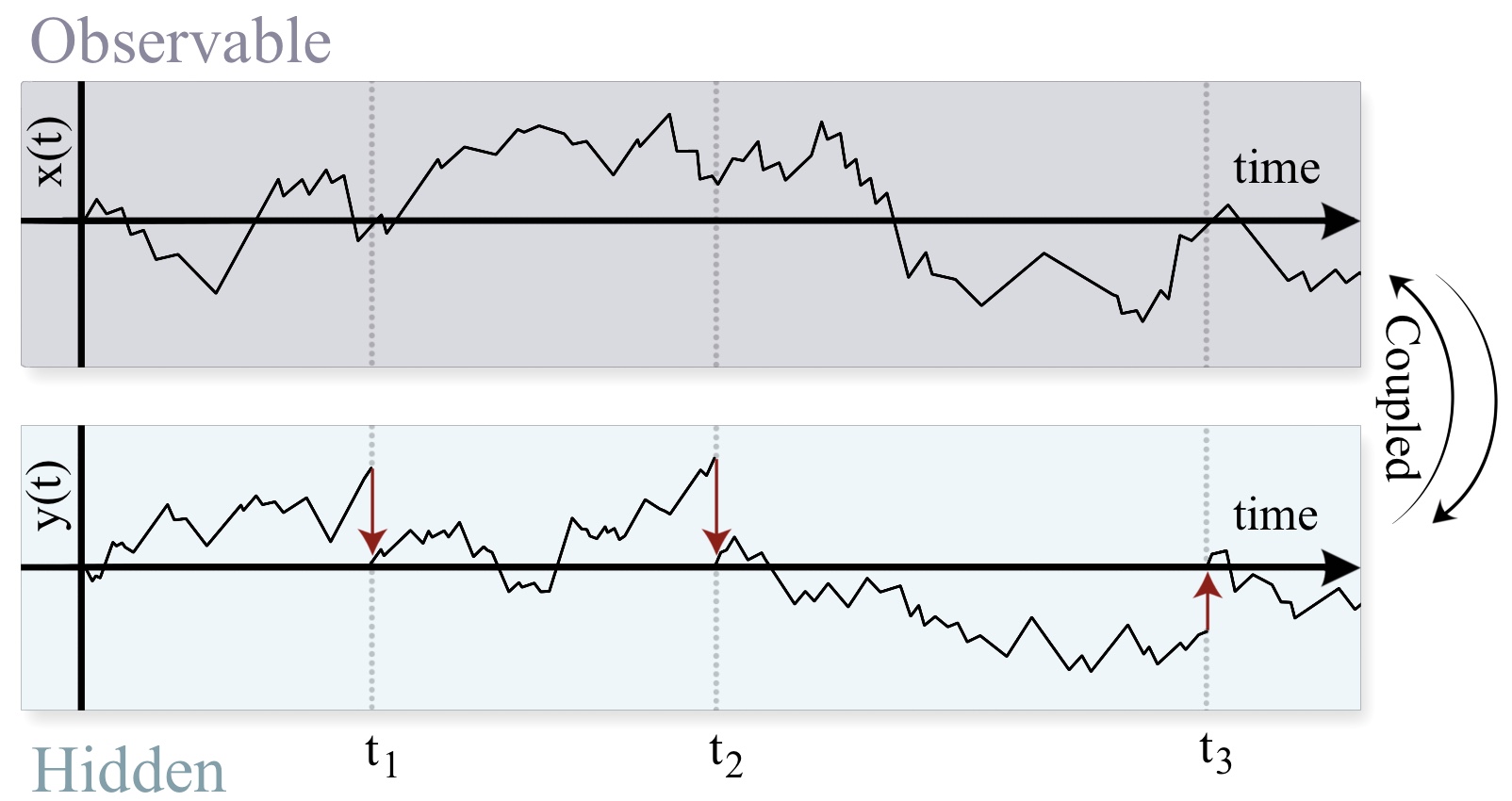}
    \caption{Sketch of the system under consideration. A set of coupled stochastic variables $(x,y)$ where only $y$ is being reset (partial resetting) at random times $\{t_i\}$.  We consider a case where the reset variable is hidden from observations, and infer indirect consequences of the resetting on the dynamics of the observable variable $x$.}
    \label{fig:sys}
\end{figure}

Many processes in nature involve degrees of freedom that evolve in seemingly stochastic ways \cite{doering2018modeling, van1992stochastic,risken1996fokker}. In many cases, large jumps in the values of an observed state variable can occur, which may drastically change the overall dynamics. Stochastic resetting is one example of large and sudden jumps where a degree of freedom is at random times reset to its initial value \cite{evans2011diffusion,evans2011optimal,evans2020review}. Over the past decade, stochastic resetting has gained much attention in the non-equilibrium statistical physics community for multiple reasons. First of all, it has been shown to optimize target search processes, with potential applications ranging from computer science to the understanding of animal foraging strategies \cite{reuveni2016optimal,pal2017first,pal2020search}. Furthermore, resetting generates \emph{non-equilibrium} steady states by trapping the system in a never-ending loop of the transient dynamical regime. Only recently has these non-equilibrium states been studied under the lens of stochastic thermodynamics, giving insights into exactly how far from thermal equilibrium such systems are \cite{fuchs2016stochastic,gupta2020work,gupta2022work,mori2023entropy,olsen2023thermodynamic,olsen2024thermodynamic}. The majority of past work is based on the dynamics and resetting of a single degree of freedom. When multiple state variables are present, with resetting only acting on a subset of these, a much richer phenomenology can occur. This paper studies such \emph{partial resetting}\footnote{One should note that the terminology \emph{partial resetting} is sometimes also used for reset processes where $x\to ax$, with $a\in(0,1)$ the strength of the resetting. Here we use the phrasing \emph{partial} in stead to refer to the resetting of parts of the set of degrees of freedom.} in a coupled two-dimensional system (see Fig. (\ref{fig:sys})), with particular focus on the case of position and velocity as is pertinent to physics.

Partial resetting has only been studied for a handful of cases in the past, to the best of our knowledge. In dimensions higher than one, one can considered resetting of only one spatial component \cite{abdoli2021stochastic}. For underdamped Brownian particles, resetting of either position alone or simultaneous reset of position and velocity has been studied \cite{gupta2019stochastic}.  Similar types of partial resets have been considered for the random acceleration processes, where both the propagator and survival probabilities have been studied \cite{singh2020random,capala2021random}.  {\color{black} Since the position of a particle can be seen as the area under
the velocity curve, connections between velocity resetting and large deviation theory can be made. In large deviation theory one often studies time-additive observables such as the area under the curve, which recently have been studied under resetting \cite{meylahn2015large,harris2017phase,den2019properties,smith2022condensation}. Velocity resetting is also natural within active matter, where resetting schemes have recently been considered where one can reset both position and direction of motion, or only one of these variables \cite{evans2018run,santra2020run,kumar2020active,mori2020universal,KSO2023}. Furthermore, run and tumble motion is in itself a velocity resetting process \cite{taylor1922diffusion,goldstein1951diffusion,lovely1975statistical}}. Recently, a similar problem where an overdamped particle in a potential is driven by a resetting noise was studied using Kesten variables, where both propagators and moments were studied \cite{gueneau2023active}. 

While much is known regarding the direct effect of resetting on a state variable, much less is known in general about the indirect effects of partial resetting in coupled systems. An example of this could be that one (or several) degrees of freedom are either not experimentally available or simply not of interest. If these un-observed, or \emph{hidden}, degrees of freedom undergo resetting, they may indirectly affect the observed degrees of freedom through cross correlations. A situation where this can occur is in the underdamped dynamics of a Brownian particle. Collisions with the environment or with a substrate may induce loss of mechanical momentum, effectively acting as resetting on the velocity variable. The position however, remains unaffected by the resetting and only changes its dynamics indirectly through its coupling to velocity. This is different from past studies of resets in underdamped dynamics where only position was reset, or both variables reset simultaneously \cite{singh2020random,gupta2019stochastic}.

Stochastic resetting of velocities can also be of relevance to various foraging strategies of animals and insects. An example is the intermittent stopping of flying foragers such as bees at flowers, whereby the velocity is reset to zero but position remains unchanged. Similar behavior, referred to as stop-and-go locomotion, is ubiquitous in the behavior of macroscopic living organisms. Here individuals intermittently stop their motion completely \cite{bartumeus2009behavioral,kramer2001behavioral}, for example in order to save energy \cite{stojan2010functions} or to scout for predators. This has for example been observed in fish motion \cite{wilson2010boldness}, in chipmunk foraging strategies  \cite{trouilloud2004head}, and in lizards climbing trees \cite{higham2011climb}. Such macroscopic systems are typically also heavily prone to inertial effects, which the framework presented in this paper incorporates by default.

In this paper we study the dynamics of a 2-variable process where only one variable undergoes resetting. { \color{black} We consider one variable to be the observed variable, which does not undergo resets, while the other variable undergoes Poissonian resetting at constant rate and is for brevity referred to as the hidden variable}. We derive explicit expressions for the Fourier-Laplace transform of the observed variables propagator, and derive from it a hierarchy of moments. We use this framework to study the dynamics of inertial particles when only velocity is being reset. We show that when velocity is reset to zero, at late times the mean squared displacement is always linear with an effective diffusivity that depends on the resetting rate. Extensions to the case where refractory periods are included after each reset is presented, in which case the late-time process may become anomalous if the refractory times are power-law distributed.

This paper is organized as follows. Section \ref{sec:prop1} discusses the propagator for two generic coupled variables when only one of them is being reset. Section \ref{sec:transport} applies these results to the case of inertial transport processes, and discusses effective transport coefficients under velocity resetting. Section \ref{sec:refr} extends these results to the case where refractory periods are included after each reset. Section \ref{sec:concl} provides a concluding discussion as well as potential outlooks.

\section{Propagator in coupled systems with resetting}\label{sec:prop1}
For the sake of simplicity we consider two coupled degrees of freedom $(x,y)$, where we observe $x$ while the variable $y$ experiences resets. Extension to multiple variables is straight-forward as long as all variables that undergo resetting do so simultaneously, in which case $y$ may be seen as a collective variable. We denote the full propagator of the system in the presence of resetting by $p_r(x,y,t|x_0,y_0)$, where $(x_0,y_0)$ are the initial conditions at $t=0$. The starting point of our analysis is the famous renewal framework for resetting processes \cite{evans2020stochastic}, which we augment to the present case of partial resetting. The last renewal equation can be used to express the propagator in terms of the propagator of the underlying, or reset-free, system $p_0(x,y,t|x_0,y_0)$ as
\begin{align}\label{eq:firstrenew}
p_r(x,y,t|x_0,y_0) &= e^{-rt} p_0(x,y,t|x_0,y_0) \nonumber\\
&+ r \int_0^t d\tau  \int dx'  dy'  p_r(x',y',t-\tau|x_0,y_0)  p_0(x,y,\tau|x',y_0) e^{-r\tau} .
\end{align}
Here the intuition is the same as in most renewal processes; the first term corresponds to realizations of the system where no resetting occurred.  These paths occur with probability $e^{-rt}$, and the system evolves with the $r=0$ propagator. The second term takes into account general trajectories (with resetting) up to the time of the last resetting $t-\tau$. The resetting takes place with probability $r d\tau$. At this instant, the system is in some arbitrary state $(x',y')$. As only $y$ is reset, the system proceeds to evolve towards $(x,y)$ from the new initial condition $(x',y_0)$. In this last time interval $\tau$ there is no resetting, again taking place with probability $e^{-r \tau}$.

Another item of note regarding Eq. (\ref{eq:firstrenew}) is the fact that all memory is deleted at the instance of reset, \emph{except for the variable $x$}. {\color{black}  Indeed, even if there were additional time dependencies coming from external effects, such as dynamical disorder from a changing environment  or diffusion with a time-dependent drift for example, even the environmental evolution must be reset for the full renewal structure assumed in Eq. (\ref{eq:firstrenew}) to be valid.} This is simply due to the fact that the free propagator which links $(x',y_0) \to (x,y)$ over the duration $(t-\tau, t)$ in Eq. (\ref{eq:firstrenew}) only depends on the duration $\tau$. Generally, time-heterogeneity can break the renewal structure one often desires when working with resetting processes, and often it is assumed that any time-dependent parameters or annealed disorder is simultaneously reset with the system variables to make the system fully renewing \cite{bodrova2019scaled,bressloff2020switching,santra2020run}. In the case of overdamped scaled Brownian motion, where temperature either grows or decreases in time, \emph{non-renewal} resetting has been studied recently, where the time dependence of the temperature is allowed to persist through a resetting event \cite{bodrova2019nonrenewal}. 

We will be interested in the the marginalized propagator of the observable variable $x_t$, which we denote 
\begin{equation}
    \wp_r(x,t|x_0,y_0) \equiv \int dy p_r(x,y,t|x_0,y_0).
\end{equation} 
and similarly for the process without resets. The indirect effect of resetting on the variable $x$ can then be obtained by integrating over $y$ in the above renewal equation:
\begin{align}
 \wp_r(x,t|x_0,y_0) &= e^{-rt} \wp_0(x,t|x_0,y_0)\\
 &+ r \int_0^t d\tau e^{-r\tau} \int dx'  \wp_r(x',t-\tau|x_0,y_0) \wp_0(x,\tau|x',y_0).\nonumber 
\end{align}
To proceed we restrict our attention to a class of spatially homogeneous systems, whereby the underlying $(r=0)$ propagator satisfies 
\begin{equation}
    \wp_0(x,\tau|x',y_0) = \wp_0(x-x',\tau|y_0).
\end{equation}
This casts the renewal equation of a convolution form, and we may readily apply a Fourier transform in space and a Laplace transform in time to solve the equation. {\color{black} We use the conventions
\begin{align}
    \mathcal{L}_s[f(t)] &\equiv \tilde f(s) = \int_0^\infty dt e^{- s t} f(t),\\
    \mathcal{F}_s[g(x)] &\equiv \hat g(k) = \int_0^\infty dt e^{-i k x } g(x.)
\end{align}
Applying a Fourier-Laplace transform to the above, we find}

\begin{align}\label{eq:main}
   \hat{\tilde{\wp}}_r  (k,s |x_0,y_0)& = \frac{ \hat{\tilde{\wp}}_0 (k,s +r |x_0,y_0)}{ 1 - r \hat{\tilde{\wp}}_0 (k,s +r |y_0)}.
\end{align}
This expresses the marginalized propagator of the observable variable $x$ under resetting of the non-observable $y$ to the corresponding propagator without resetting. Note that only the numerator is conditioned on both initial conditions, while the term in the denominator should be taken with $x_0=0$ as a consequence of the homogeneity assumption. {\color{black} Propagators of this form were recently studied in detail for fractional Brownian motion in Ref. \cite{smith2022condensation}. Here we use Eq. (\ref{eq:main}) to derive a hierarchy of moments from which exact expressions for the coefficients governing late-time scaling is provided. We later extend these results to also allow refractory periods of arbitrary durations after each reset.}

\subsection{Hierarchy of moments}
Inverting the above solution in  Eq. (\ref{eq:main}) exactly can be arduous in many cases, as the marginalized propagator even in simple coupled systems have complex Fourier-Laplace transforms. To make further progress, we derive from it a general expression for the moments of the process  $x_t$. First, note that moments can be computed from Fourier transforms of the probability density as 

\begin{align}\label{eq:mom1}
     {\langle  x^n | x_0,y_0 \rangle_t} = \frac{\partial^n}{\partial (ik)^n} \hat{{ \wp }}_r(k,t|x_0,y_0) \bigg |_{k=0}.
\end{align}
As we have the Fourier-Laplace transform, we will in our case find 
\begin{align}\label{eq:mom2}
   \widetilde{\langle  x^n | x_0,y_0 \rangle}_s= \frac{\partial^n}{\partial (ik)^n} \hat{\tilde{ \wp }}_r(k,s|x_0,y_0)\bigg |_{k=0},
\end{align}
where $ \widetilde{\langle  x^n | x_0,y_0 \rangle}_s=  \mathcal{L}_{t\to s} \left [ {\langle  x^n | x_0,y_0 \rangle_t} \right]$ denotes Laplace transforms.  In order to obtain expressions for the moments from Eq. (\ref{eq:main}), we first re-write it as 
\begin{align}
   \hat{\tilde{\wp}}_r  (k,s |x_0,y_0) [1 - r \hat{\tilde{\wp}}_0 (k,s +r |y_0)] =  \hat{\tilde{\wp}}_0 (k,s +r |x_0,y_0)
\end{align}
simply to avoid having to deal with quotients. Using the generalized product rule for higher-order derivatives, we have 
\begin{align}
 \frac{\partial^n}{\partial (ik)^n}   \hat{\tilde{\wp}}_0 (k,s +r |x_0,y_0) &=   \sum_{\ell = 0}^n {n \choose \ell} \frac{\partial^{(n-\ell)}}{\partial (ik)^{(n-\ell)}} \hat{\tilde{\wp}}_r  (k,s |x_0,y_0) \nonumber \\
  &\times \frac{\partial^\ell}{\partial (ik)^\ell}  [1 - r \hat{\tilde{\wp}}_0 (k,s +r |y_0)] .
\end{align}
Setting $k=0$ and using the expression for the moments, we have 
\begin{align}
  \widetilde{\langle  x^{n}|x_0,y_0\rangle}_{s+r}^{(0)}  =   \sum_{\ell = 0}^n {n \choose \ell} \widetilde{\langle  x^{n-\ell} |x_0,y_0\rangle}_s \: \: [ \delta_{\ell,0} - r\widetilde{\langle  x^{\ell} |y_0 \rangle}_{s+r}^{(0)}]  ,
\end{align}
where the superscript ${(0)} $ denotes moments for the underlying process with $r=0$. Rearranging gives 

\begin{align}\label{eq:main2}
  \widetilde{\langle  x^{n} |x_0,y_0\rangle}_s   & =  \frac{s+r}{s} \widetilde{\langle  x^{n}|x_0,y_0\rangle}_{s+r}^{(0)} +\frac{s+r}{s} \left[   r  \sum_{\ell = 1}^n {n \choose \ell} \widetilde{\langle  x^{n-\ell} |x_0,y_0\rangle}_s  \widetilde{\langle  x^{\ell} |y_0 \rangle}_{s+r}^{(0)}   \right]  .
\end{align}

This recursive relation can be used to iteratively construct any moment of the process $x_t$ given lower-order moments and the moments of the $r=0$ case, which are assumed to be known. We emphasize that this result can be useful when the Fourier-Laplace transform of the propagator itself is hard to obtain, while if $ \hat{\tilde{p}}_0 (k,s  |x_0,y_0)$ is known one can simply expand Eq. (\ref{eq:main}) in powers of $k$ to identify the moments. We also note that when $r=0$, the above equation reduces to $\langle  x^{n} |x_0,y_0\rangle_s=\langle  x^{n} |x_0,y_0\rangle_s^{(0)}$ as it should.

\section{Transport processes: crossover from anomalous to normal diffusion}\label{sec:transport}

For transport processes the relevant physical variables are often position {\color{black} $x_t$} and velocity {\color{black} $v_t$} of a particle, {\color{black} satisfying $\dot x_t = v_t$}. Here we consider the effect of velocity resetting on spatial transport by using Eq. (\ref{eq:main2}). This is of relevance to a wide range of systems, for example for particles with inelastic collisions with an environment or substrate, or in foraging processes where animals intermittently stop (eg. to collect nutrients, or scout for predators) before re-starting their motion from zero velocity but unchanged position. From the hierarchy in  Eq. (\ref{eq:main2}) we derive general expressions for the effective drift and diffusivity, and consider in more detail the case where the underlying process shows anomalous diffusion.

\subsection{Effective transport coefficients}

To characterize transport, we are interested in the first two spatial moments

\begin{align}
       \widetilde {\langle x| x_0,v_0\rangle}_s &= \frac{s+r}{s}\left [ \widetilde{\langle x| x_0,v_0\rangle}_{s+r}^{(0)}  +  \frac{r}{s}  \widetilde{\langle x| v_0\rangle}_{s+r}^{(0)}    \right],  \\
    \widetilde{\langle x^2| x_0,v_0\rangle}_s & = \frac{s+r}{s}\widetilde{\langle x^2| x_0,v_0\rangle}_{s+r}^{(0)} + \frac{s+r}{s}  \frac{r}{s}\widetilde{\langle x^2| v_0\rangle}_{s+r}^{(0)} \nonumber \\
    & + \frac{s+r}{s} \left [    2 r \widetilde{\langle x| x_0,v_0\rangle}_{s} \widetilde{\langle x| v_0\rangle}_{s+r}^{(0)}     \right] ,
\end{align}
which contain information regarding effective drift and dispersion of the spatial variable. Without loss of generality, let us consider $x_0 =0$. Then the first moment can be written compactly as 
\begin{align}\label{eq:fst}
\widetilde{\langle x| 0,v_0\rangle}_s = \left( \frac{s+r}{s} \right)^2
      \widetilde{\langle x| 0,v_0\rangle}_{s+r}^{(0)}.
\end{align}
At late times, corresponding to small values of $s$, the $s^{-2}$ pole dominates, giving rise to a linear growth in time. The inverse Laplace transform can be calculated in terms of residues as
\begin{align}
    \langle x| 0,v_0\rangle_t &= \mathcal{L}^{-1}_{s\to t}\left \{\left( \frac{s+r}{s} \right)^2  \widetilde{\langle x| 0,v_0\rangle}_{s+r}^{(0)} \right\}  \\
    & = \sum_\text{poles $\{s_i\}$} \text{Res}_i\left[ \left( \frac{s+r}{s} \right)^2 \widetilde{ \langle x| 0, v_0\rangle}_{s+r}^{(0)} e^{st}\right],\nonumber
\end{align}
where the poles $\{s_i\}$ are those of Eq. (\ref{eq:fst}). The poles at non-zero values of $s$ gives rise to exponentially decaying terms. Hence the late-time behavior is extracted from the pole at zero, in which case the residue reads
\begin{align}
      \langle x| 0,v_0\rangle_t &= \lim_{s\to 0}\partial_s \left[ (s+r)^2  \widetilde{\langle x| 0,v_0\rangle}_{s+r}^{(0)} e^{st}\right]= r^2  \widetilde{\langle x| 0,v_0\rangle}_{r}^{(0)} t  + ...
\end{align}
where the terms $+...$ represent terms that are either approaching a constant or vanishing at late times. The effective drift at late times is then calculated as
\begin{equation}\label{eq:drift}
    V_\text{eff} \equiv \lim_{t\to \infty} \frac{\langle x| 0,v_0\rangle_{t}}{ t} = r^2 \widetilde{\langle x| 0,v_0\rangle}_{r}^{(0)}.
\end{equation}
Hence velocity resetting to a non-zero value of velocity $v_0$, which typically results in non-zero $\langle x| 0,v_0\rangle_t^{(0)}$, will result in a drift at late times, as is expected \footnote{It should be noted that one may in principle reset velocity to a non-zero value $v_0$ and yet obtain a vanishing mean $\langle x| 0,v_0\rangle_t^{(0)}$. For example, in the presence on a constant external drift one could reset to a velocity that exactly opposes the drift.}. This can be seen as a rectification effect due to velocity resetting. {\color{black} A physical intuition of Eq. (\ref{eq:drift}) is obtained by noting that the late-time behavior of the mean position can be written 
\begin{equation}
     \langle x| 0,v_0\rangle_t \simeq V_\text{eff} t =  (r t) \int dt r e^{-r t}  \langle x| 0,v_0\rangle_t^{(0)}
\end{equation}
which is nothing but the mean number of resets $\overline{n} = rt$ in time $t$, times the mean step length during a inter-reset epoch.
}

In the remainder of this section, {\color{black} we consider for simplicity symmetric processes with vanishing odd moments so that no drift is present $V_\text{eff} =0$, and set $v_0 =0$.} Proceeding similarly as for the effective drift, we can derive an expression for the effective diffusivity. The second moment in Laplace space reads
\begin{align}\label{eq:simp_x2}
    \widetilde{\langle x^2| 0,0\rangle}_s =& \left( \frac{s+r}{s} \right)^2  \widetilde{\langle x^2| 0,0\rangle}_{s+r}^{(0)} .
\end{align}
By the same logic, the $s^{-2}$ pole gives rise to a linearly growing mean-squared-displacement. Again, the real-time mean-squared-displacement can be expressed as a sum over poles $s_i$ of Eq. (\ref{eq:simp_x2})
\begin{align}
    \langle x^2| 0,0\rangle_t &= \mathcal{L}^{-1}_{s\to t}\left \{\left( \frac{s+r}{s} \right)^2  \widetilde{\langle x^2| 0,0\rangle}_{s+r}^{(0)} \right\}  \\
    & = \sum_\text{poles $\{s_i\}$} \text{Res}_i\left[ \left( \frac{s+r}{s} \right)^2 \widetilde{ \langle x^2| 0,0\rangle}_{s+r}^{(0)} e^{st}\right].\nonumber
\end{align}
The dominant late-time behavior once again comes from the second-order pole at $s=0$, resulting in
\begin{align}\label{eq:interm}
    \langle x^2| 0,0\rangle_t &= \lim_{s\to 0}\partial_s \left[ (s+r)^2  \widetilde{\langle x^2| 0,0\rangle}_{s+r}^{(0)} e^{st}\right] = r^2 \widetilde{\langle x^2| 0,0\rangle}_{ r}^{(0)} t  +...
\end{align}
Hence the effective diffusivity can be calculated as 
\begin{equation}\label{eq:gendeff}
    D_\text{eff} \equiv \lim_{t\to \infty} \frac{\langle x^2| 0,0\rangle_{t}}{2 t} = \frac{r^2}{2} \widetilde{\langle x^2| 0,0\rangle}_{r}^{(0)}.
\end{equation}
{ \color{black} As for the effective drift, a natural physical interpretation of this in terms of the asymptotic behavior of the mean squared displacement can be obtained by noticing that 
\begin{equation}
    \langle x^2| 0,0\rangle_{t} \simeq 2 D_\text{eff} t =  rt \int_0^\infty dt re^{-rt}  \langle x^2| 0,0\rangle_{t}^{(0)}
\end{equation}
which is the mean number of resets $\overline{n}$ times the mean growth of the second moment during a inter-reset epoch.} We stress here the generality of this result; as long as the system considered is spatially homogeneous, and the full renewal structure of Eq. (\ref{eq:firstrenew}) is satisfied, the process under velocity resets will display normal diffusion with the above effective diffusivity, even if the reset-free system shows anomalous diffusion. We discuss this further in later sections. \\

\begin{figure}
    \centering
    \includegraphics[width = 9.7cm]{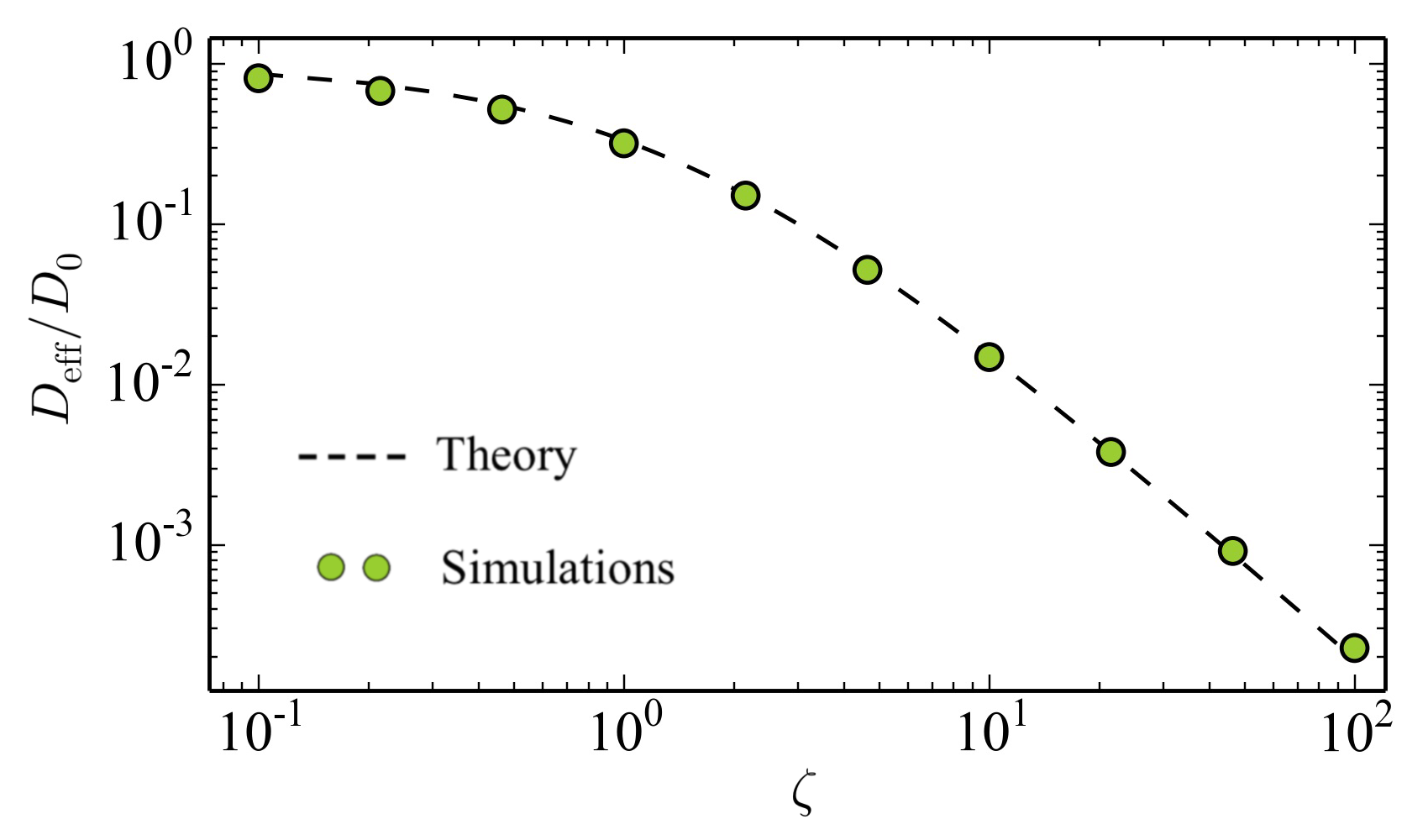}
    \caption{Effective diffusivity for an underdamped Brownian motion undergoing velocity resetting, as a function of re-scaled resetting rate $\zeta = r/\gamma$. Dashed black lines show exact theory (Eq. (\ref{eq:UBPdeff})) while points show simulated data, using $D_0 = \gamma = 1$.}
    \label{fig:UBP}
\end{figure}

\textbf{Underdamped Brownian motion:} As a simple application of the above formula, we consider velocity resetting for an underdamped Brownian particle.  This can be seen as a minimal model for a particle moving through a complex environment, where intermittently the particle collides with the environment and looses its momentum \cite{pierce2022advection}. The particle obeys the coupled equations 
\begin{align}
     d x_\tau &= v_\tau dt, \label{eq:udbm1} \\
     d v_\tau &= -\gamma v_\tau dt + \sqrt{2k_B T \gamma} dW , \label{eq:udbm2}
\end{align}
where $dW$ is an increment in the Wiener process and $\gamma$ the friction coefficient. Here we have set mass to unity. The marginalized position distribution reads \cite{gupta2019stochastic}
\begin{equation}
    \wp_0(x,t | x_0,v_0) = \frac{1}{\sqrt{2 \pi \Sigma^2_t}} \exp \left(- \frac{(x-\langle x_t\rangle)^2}{2\Sigma^2_t}\right),
\end{equation}
where the variance and mean of the density reads
\begin{align}
    \Sigma^2_t &= D_0 \tau_\gamma \left(\frac{2 t}{\tau_\gamma} + 4 e^{-t/\tau_\gamma}- e^{-2 t /\tau_\gamma} - 3 \right),\\
    \langle x_t\rangle &= x_0 + v_0 \tau_\gamma \left(1 - e^{- t /\tau_\gamma}\right),
\end{align}
with $\tau_\gamma = 1/\gamma$ being the inertial timescale and $D_0 = k_B T / \gamma$. From this one can easily calculate the second moment in Laplace space, resulting in
\begin{equation}
   \widetilde{  \langle x^2| 0,0 \rangle}_s^{(0)} = -\frac{3 D_0}{\gamma s}+\frac{4 D_0}{\gamma  (\gamma +s)}-\frac{D_0}{\gamma  (2 \gamma
   +s)} +\frac{2 D_0}{s^2},
\end{equation}
where we set $x_0 = v_0 = 0$, so that $V_\text{eff} = 0$. Using Eq. (\ref{eq:gendeff}) one immediately finds 
\begin{equation}
   \frac{D_\textnormal{eff}}{D_0} = \left( 1 + \zeta  \:\frac{3 + \zeta}{2 } \right)^{-1} .\label{eq:UBPdeff}
\end{equation}
Here we introduced the dimensionless variable $\zeta = r/\gamma$. As expected, $D_\text{eff} = D_0$ is obtained when $\zeta = 0$, {\color{black} which can be achieved either by $r=0$ or in the high-friction limit $\gamma \to \infty$. The latter case highlights the fact that for Brownian particles a non-trivial effective diffusivity only occurs in the presence of inertial effects, whereby the particle must accelerate after each reset.} We also see that velocity resetting suppresses spatial transport, and $D_\textnormal{eff}=0$ as $r\to \infty$. This is verified using numerical simulations in Fig. (\ref{fig:UBP}).

\subsection{When the underlying process is anomalous}
The above results have interesting implications for transport processes in $(x,v)$ that would be anomalous in the absence of resetting. {\color{black} Anomalous diffusion processes with resetting have been studied recently for overdamped systems in the context of ergodicity breaking and restoration  \cite{wang2022restoring,vinod2022time,vinod2022nonergodicity,liang2023anomalous}. Here we investigate the effects of velocity resetting in underdamped anomalous systems.}

For anomalous diffusion processes, the underlying process satisfies in the simplest case
\begin{align}\label{eq:anom}
     \langle x^2| 0,0\rangle_{t}^{(0)} = D_0 t^\alpha,
\end{align}
with diffusion exponent $\alpha$ taking any positive real value. Such processes are present in a wide range of systems, with classical examples being transport of Brownian particles in complex environments such as fractals or media with power-law friction \cite{havlin1987diffusion,bouchaud1990anomalous,ben2000diffusion,sokolov2012models,olsen2019geometric,olsen2020diffusion,olsen2021active}. Anomalous diffusion has also been observed in granular systems, such as in the velocity profile of fluid-drive silo discharge \cite{morgan2021subdiffusion}, in granular gasses near a shear instability \cite{brey2015anomalous}, and in the height fluctuations in graphene \cite{granato2022dynamic}. In such cases, one often expects subdiffusive processes $\alpha < 1$, while there are also cases where superdiffusion with $\alpha > 1$ occurs, such as in L\'{e}vy flights, random acceleration processes, tracers in turbulent flow, active particles with time-dependent self-propulsion forces, and in diffusion with density-dependent diffusivity  \cite{dubkov2008levy,richardson1926atmospheric,burkhardt2014first,hansen2020anomalous,babel2014swimming,flekkoy2021hyperballistic}. For a review of theoretical models of anomalous diffusion, see for example \cite{metzler2014anomalous}. While most of these studies consider particles in the overdamped limit, with no coupled variables, many models can be extended to the underdamped case and studied under the framework presented here.

The Laplace transform in time of Eq. (\ref{eq:anom}) reads $ \langle x^2| 0,0\rangle_{s}^{(0)} = D_0 s^{-(1 + \alpha)} \Gamma(1 + \alpha)$. Using Eq. (\ref{eq:simp_x2}) this leads directly to 
\begin{align}
    \widetilde{\langle x^2| 0,0\rangle}_s =\frac{D_0 \Gamma (\alpha +1) (r+s)^{1-\alpha }}{s^2}.
\end{align}
Inverting gives the full time-dependence of the mean-squared-displacement
\begin{align}\label{eq:fullmsd}
  &  \langle x^2| x_0,v_0\rangle_t =(\alpha -1) D_0 r^{-\alpha } (\alpha  \Gamma (\alpha ,r t)-\Gamma (\alpha +1))\nonumber\\
    & + (\alpha -1) \alpha  D_0  r^{1-\alpha } (\Gamma (\alpha -1)-\Gamma (\alpha -1,r t)) t, 
\end{align}
where $\Gamma(x,z)$ is the  (upper) incomplete Gamma function. At early times, as resetting has not yet had time to occur, the mean-squared-displacement behaves as in the underlying system. Generally, the short-time behavior of the mean squared displacement can be obtained by looking at large values of its Laplace variable. For $s\gg  r$ in Eq. (\ref{eq:simp_x2}) we see that $\langle x^2|0,0\rangle_s = \langle x^2|0,0\rangle_s^{(0)}$. Hence in this case we have $\langle x^2| 0,0\rangle_t  = D_0 t^\alpha $. At late times, we see that a linear growth takes over, in agreement with what we  predicted more generally (see Eqs. (\ref{eq:interm}) and (\ref{eq:gendeff})). The effective diffusivity can be calculated either from Eq. (\ref{eq:gendeff}) or directly from Eq. (\ref{eq:fullmsd}), yielding
\begin{equation}\label{eq:deffex}
    D_\textnormal{eff} \equiv \lim_{t\to \infty} \frac { \langle x^2| 0, 0\rangle_t}{2 t} = \frac{D_0 \Gamma (\alpha +1)  }{2 r^{\alpha-1}}  .
\end{equation}
We see that anomalous diffusion processes with diffusion exponent $\alpha$ become normal under velocity resetting, with an effective diffusion coefficient scaling with resetting rate as $D_\textnormal{eff} \sim r^{1-\alpha}$. For superdiffusive processes, the effective diffusivity decreases as a function of resetting rate, while for subdiffusive processes it grows with increasing resetting rate. {\color{black} The fact that subdiffusive processes can enhance their diffusivity by resetting originates in the full renewal structure of Eq. (\ref{eq:firstrenew}). Since subdiffusion often occurs for processes that slow down substantially over time, such as for single-file diffusion \cite{wei2000single,kollmann2003single}, restarting to a state of higher motility can be beneficial.}  The crossover time $t_*$ where the linear growth starts to dominate over the anomalous growth can be identified by matching the late-time regime with effective diffusivity given by Eq.(\ref{eq:deffex}) with the early-time growth $D_0 t^\alpha$, resulting in 
\begin{equation}\label{eq:tcross}
    t_* = \left[\frac{\Gamma(\alpha+1)}{2} \right]^{\frac{1}{\alpha-1}} r^{-1}.
\end{equation}
Since the resetting timescale $r^{-1}$ determines when the mean squared displacement should cross over to linear growth, this proportionality is sensible.

\begin{figure*}
    \centering
    \includegraphics[width = \textwidth]{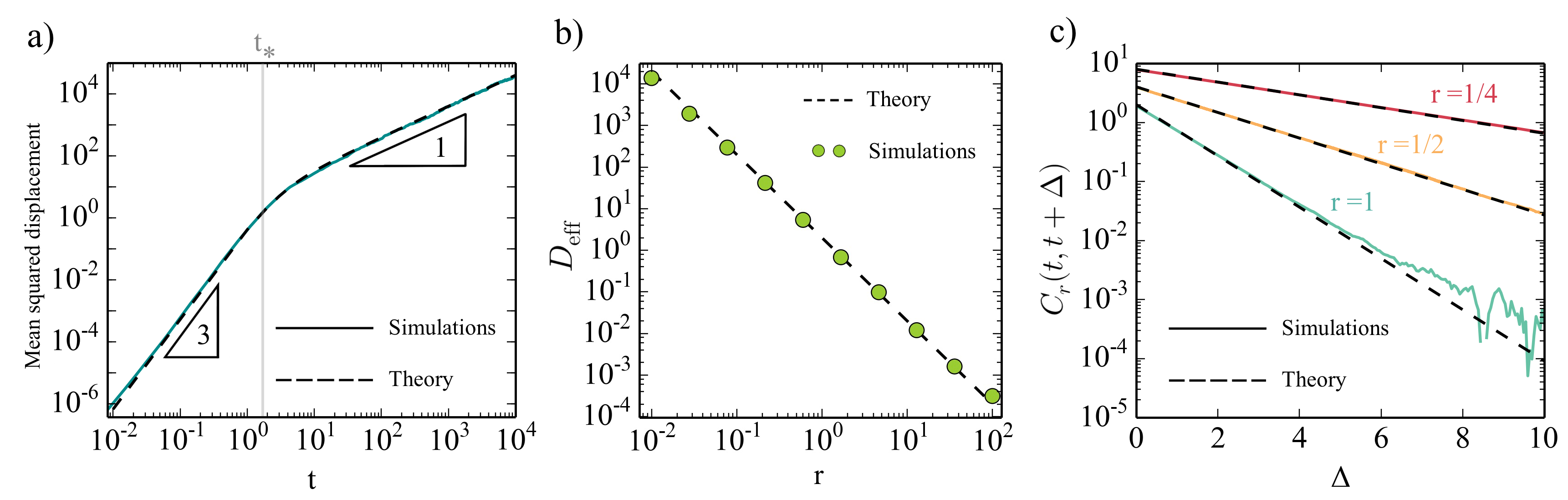}
    \caption{Comparisons of theory (dashed black lines) and simulated data for a random acceleration process. a) Mean squared displacement shows a clear crossover from anomalous $\alpha =3$ diffusion to normal diffusion with $\alpha = 1$ at times larger than the crossover time $t_*$(grey vertical line). b) Effective diffusivity at late times scales as $D_\text{eff} \sim r^{-2}$, as predicted by Eq. (\ref{eq:deffex}). c) Velocity auto-correlation functions showing exponential tempering. Solid colored lines are simulations, while black dashed line corresponds to Eq. (\ref{eq:correx}). Resetting rate is set to unity unless stated otherwise. }
    \label{fig:RAP}
\end{figure*}

\textbf{Random acceleration process:} A concrete example of the above could be the random acceleration process \cite{burkhardt2014first}
\begin{align}
    \dot x &= v,\\
    \dot v &= \sqrt{2}\eta(t),
\end{align}
where $\eta(t)$ is a Gaussian white noise with correlator $\langle \eta(t_1) \eta(t_2)\rangle = \delta(t_1-t_2)$, and where we set the noise strength to unity for simplicity. The marginalized process $\wp_0(x,t|x_0,v_0)$ is known to have the Gaussian propagator \cite{burkhardt2014first,singh2020random}
\begin{equation}
    \wp_0(x,t|0,0) = \sqrt{\frac{3}{4\pi t^3}} \exp\left(- \frac{3}{4 t^3} x^2 \right),
\end{equation}
with a mean-squared-displacement that grows as a cubic in time $\langle x^2 |0,0\rangle_t^{(0)} = \frac{2}{3}t^3$, i.e. $\alpha = 3$ and $D_0 = 2/3$ in the above  calculation.  The effective diffusivity after resetting reads $D_\text{eff} = 2/r^2$, with a crossover time $t_* = \sqrt{3}/r$. {\color{black} This result was also reported in Ref. \cite{smith2022condensation} obtained by other methods.} Here the full mean-squared-displacement can easily be obtained from Eq. (\ref{eq:fullmsd}), which we show in Fig. (\ref{fig:RAP} a), where a clear crossover from anomalous $\alpha = 3$ to normal $\alpha = 1$ is seen. The predicted $D_\text{eff}\sim r^{-2}$ scaling also matches perfectly with simulated data as seen in Fig. (\ref{fig:RAP} b).

\subsection{Tempered correlation functions}
The emergence of normal diffusion independently of the anomalous nature of the underlying process can be understood as a consequence of resetting-induced tempering of the velocity auto-correlation functions. It is known that systems with exponential (or sufficiently strong power-law)  cutoffs, displays a crossover from anomalous to normal diffusion at late times \cite{molina2018crossover}. 

For transport processes, the position is coupled to velocity simply through $\dot x = v$, implying that the mean-squared displacement in general comes from the temporal behavior of the velocity correlation functions
\begin{equation}\label{eq:kubo}
    \langle (x-x_0)^2\rangle_t= 2 \int_0^t d\tau_2 \int_0^{\tau_2}d \tau_1 C_r(\tau_1,\tau_2),
\end{equation}
where we introduced $C_r(\tau_1,\tau_2) =  \langle v_{\tau_1}v_{\tau_2} \rangle $. We will by $ C_0(\tau_1,\tau_2)$ denote the velocity auto-correlations in the absence of resetting. For correlations that decay sufficiently fast, the system exhibits normal diffusion and the above equation can be turned into a  Green-Kubo relation for the diffusion coefficient. If the correlation functions decay too slowly, or even grow in time, one expects anomalous diffusion.

Similarly to the probability density,  correlation functions are known to satisfy a renewal equation of the type \cite{majumdar2018spectral,evans2020stochastic}
\begin{align}\label{eq:rencorr}
    C_r(t_1,t_2) &= e^{- r t_2} C_0(t_1,t_2)  \\
    &+ r e^{-r(t_2-t_1)}\int_0^{t_1} d\tau e^{-r\tau} C_0(\tau,t_2-t_1+\tau). \nonumber
\end{align}
Here we have relied on the assumption that position is coupled to velocity, but not vice-versa, allowing a simple renewal equation for the velocity correlations. This is normal for most models of Brownian motion, and has been utilized in the past where considering underdamped Brownian motion under partial resetting of position alone \cite{gupta2019stochastic}. Already we can see that the inclusion of resetting tempers the correlations by including an exponential cutoff at a characteristic time $r^{-1}$ already in the first term. 

Returning again to a concrete example where the underlying process is anomalous of the type Eq. (\ref{eq:anom}), we consider correlation functions of the form
\begin{equation}\label{eq:vc}
    C_0(t_1,t_2) = \frac{\alpha(\alpha-1) D_0}{2} \text{min}\{t_1,t_2\}^{\alpha-2}.
\end{equation}
{\color{black} This correlation function is chosen so that by applying  Eq. (\ref{eq:kubo}) gives rise to a mean-squared-displacement growing anomalously in time with exponent $\alpha$.} Under resetting, the new correlator Eq. (\ref{eq:rencorr}) reads
\begin{equation}\label{eq:correx}
    C_r(t_1,t_2) =  \frac{ \alpha (\alpha -1)   D_0 }{2}e^{-r ({t_2}-{t_1})}  \left(r^{2-\alpha} \left(\Gamma
   (\alpha-1)-\Gamma \left(\alpha-1,r t_1\right)\right)+t_1^{\alpha-2} e^{-r t_1}\right) , \nonumber
\end{equation}
where we assumed $t_2\geq t_1$. To clearly see the tempering, consider $ C_r(t,t + \Delta) $ with $\Delta\geq 0$ a lag variable. For $t \gg r^{-1}$ we then find 

\begin{equation}
   C_r(t,t + \Delta) \simeq  \frac{\Gamma
   (\alpha+1)   D_0 r^{2-\alpha} }{2}   e^{-r \Delta} ,
\end{equation}
which shows a clear cutoff scale $r^{-1}$. This exponential behavior is shown for the random acceleration process in Fig. (\ref{fig:RAP} c), where normally the velocity correlations obey Eq. (\ref{eq:vc}) with $\alpha = 3$.

\subsection{The effect of real-time crossovers}

As a final example, we consider a case where the underlying system has a crossover from one dynamical regime to another at a crossover time $t_c$. We assume 
\begin{equation}
    \langle x^2 \rangle^{(0)}_t = 2D_\alpha t^\alpha \theta(t_c-t) + 2D_\beta t^\beta \theta(t-t_c).
\end{equation}
The effective diffusivity can again be calculated from Eq. (\ref{eq:gendeff}), resulting in 
\begin{equation}
    D_\text{eff} (r) = D_\alpha r^{1-\alpha}\gamma(\alpha+1, r t_c) + D_\beta r^{1-\beta}\Gamma(\beta+1, r t_c),
\end{equation}
where $\Gamma(\cdot)$ and $\gamma(\cdot)$ are the upper and lower incomplete Gamma functions respectively. We note that while $\gamma(a,z)$ vanishes as $z\to 0$, $\Gamma(a,z)$ simply approaches $\Gamma(a)$ in this limit. A converse behavior holds for $z\to \infty$. Hence the incomplete gamma functions {\color{black}give more or less weight} to the two scaling behaviors depending on the value of the resetting rate. In particular, the effective diffusivity changes scaling behavior as a function of resetting rate $r$ at a {\color{black}crossover} value $r_c = t_c^{-1}$. In the case {\color{black}$r\ll t_c^{-1}$}, the system resets so rarely that the late-time dynamical regime always can be explored, and $D_\text{eff}(r) \sim r^{1-\beta}$. Similarly, when {\color{black}$r\gg t_c^{-1}$} typical trajectories only explore the first dynamical regime, and $D_\text{eff}(r) \sim r^{1-\alpha}$. 

The underdamped Brownian particle {\color{black} described by Eqs. (\ref{eq:udbm1}) \& (\ref{eq:udbm2})} yet again serves as a good example for a case with a crossover. In this case, it is known that for a particle initially at rest, the short-time mean squared displacement has a leading-order cubic behavior \cite{breoni2020active}
\begin{equation}
    \langle x^2 \rangle^{(0)}_t  =\frac{4 \gamma^2 D_0}{3} t^3 +\mathcal{O}(t^4) ,
\end{equation}
while at late times it crosses over to normal diffusion $\langle x^2 \rangle^{(0)}_t \sim t$. Hence we expect an effective diffusivity that first is constant before decaying as $r^{-2}$. This is indeed what is observed in Fig. (\ref{fig:UBP}), or equivalent in Eq. (\ref{eq:UBPdeff}).

\vspace{0.4cm}

\begin{figure*}[t]
    \centering
    \includegraphics[width = \textwidth]{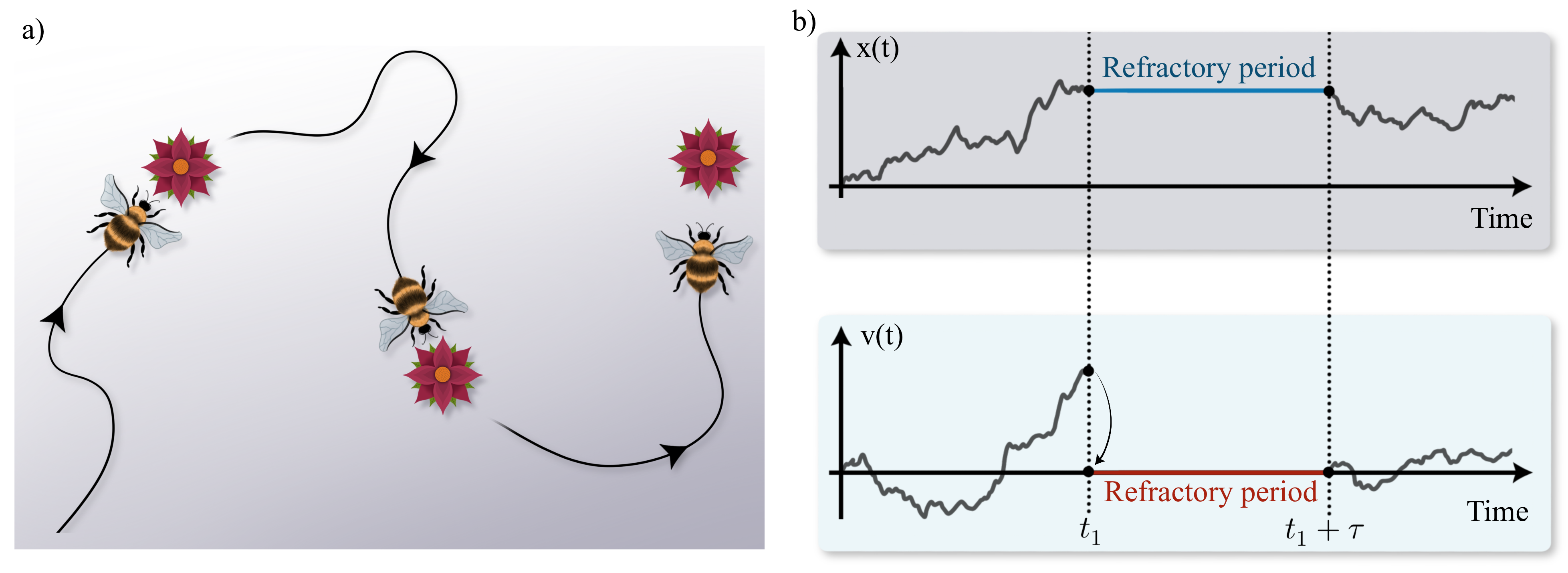}
    \caption{Sketch of the system we consider. a) Typical dynamics of a forager consists in exploration phases interrupted by refractory periods of no motion. b) Sketch of one dimension time-series for position and velocity, showing a refractory period of duration $\tau$ following a resetting at time $t_1$. Here we assume that velocity is reset to zero, so that no motion occurs during the refractory period.}
    \label{fig:sketch2}
\end{figure*}

\section{Effects of refractory periods}\label{sec:refr}

In the case of inertial foragers that perform stop-and-go locomotion, velocity resets to $v_0=0$ are often followed by inactive periods (see Fig. (\ref{fig:sketch2})). In the context of resetting, such idle times after resetting events are referred to as refractory times \cite{evans2018effects,garcia2023stochastic}. In overdamped systems with a single degree of freedom, anomalous diffusion behavior has been observed for particular choices of refractory and inter-reset statistics \cite{maso2019stochastic}. In this section we include the effect of such refractory times in underdamped systems, and show how normal Poissonian resets of velocity can also lead to anomalous diffusion.

\subsection{Exact propagator}
As for the case of velocity resetting without refractory times, we start by investigating the renewal equation. The first renewal equation for the propagator can be written

\begin{align}
   & p_r(x,v,t|x_0,v_0) = e^{-rt}p_0(x,v,t|x_0,v_0)\nonumber\\
    & + r \int_0^t \rmd t_1 e^{-r t_1}\int_0^{t-t_1}\rmd\tau W(\tau)\int \rmd y \rmd u p_0(y,u,t_1|x_0,v_0) p_r(x,v,t-t_1-\tau|y,v_0) \nonumber \\
    & + r\int_0^t \rmd t_1 e^{-r t_1} \int_{t-t_1}^\infty \rmd \tau W(\tau) \int \rmd y \rmd u p_0(y,u,t_1|x_0,v_0)\delta(v-v_0)\delta(x-y).
\end{align}
Here the first two terms have  the same interpretation as before, corresponding respectively to paths without resets and paths with resets. The only difference is the inclusion of a refractory time $\tau$ after the first reset event. The third term corresponds to paths that at time $t$ end in the refractory phase.

As before, we will assume spatial homogeneity, which means we can without loss of generality set $x_0 =0$. To ease notation, as before, we suppress this from the propagator so that $p_i(x,v,t|x_0 = 0 ,v_0) = p_i(x,v,t|v_0)$, with $i = r,0$. Integrating the above first renewal equation over velocity $v$ we find the position distribution
\begin{align}
    \wp_r(x,t|v_0) &= e^{-rt}\wp_0(x,t|v_0)\nonumber\\
    & + r \int_0^t \rmd t_1 e^{-r t_1}\int_0^{t-t_1}\rmd\tau W(\tau)\int \rmd y \wp_0(y,t_1|v_0) \wp_r(x,t-t_1-\tau|y,v_0) \nonumber \\
    & + r\int_0^t \rmd t_1 e^{-r t_1} \int_{t-t_1}^\infty \rmd \tau W(\tau)  \wp_0(x,t_1|v_0) .
\end{align}
Invoking spatial homogeneity we write $\wp_r(x,t-t_1-\tau|y,v_0) = \wp_r(x-y,t-t_1-\tau|v_0)$. Taking a Fourier transform in space, we obtain 
\begin{align}
    \hat\wp_r(k,t|v_0) &= e^{-rt}\hat\wp_0(k ,t|v_0)\nonumber\\
    & + r \int_0^t \rmd t_1 e^{-r t_1}  \hat \wp_0(k,t_1|v_0)  \int_0^{t-t_1}\rmd\tau W(\tau) \hat\wp_r(k,t-t_1-\tau|v_0) \nonumber \\
    & + r\int_0^t \rmd t_1 e^{-r t_1}  \hat \wp_0(k,t_1|v_0) \int_{t-t_1}^\infty \rmd \tau W(\tau) .\label{eq:firstrenewal}
\end{align}
Laplace transforming Eq. (\ref{eq:firstrenewal})  results in 

\begin{align}
    \hat{\tilde{\wp}}_r(k,s|v_0) &= \hat{\tilde{\wp}}_0(k,s+r|v_0) + r \tilde{W}(s) \hat{\tilde{\wp}}_0(k,s+r|v_0)\hat{\tilde{\wp}}_r(k,s|v_0) \nonumber  \\
    &+ r \frac{\hat{\tilde{\wp}}_0(k,s+r|v_0)}{s}\left[ 1-\tilde {W}(s)\right].
\end{align}
Solving for the propagator in the presence of resetting, we get the closed form expression
\begin{equation}\label{eq:exact}
    \hat{\tilde{\wp}}_r(k,s|v_0) = \frac{\hat{\tilde{\wp}}_0(k,s+r|v_0)}{1 - r \tilde{W}(s) \hat{\tilde{\wp}}_0(k,s+r|v_0)}\left( 1 - \frac{r}{s} [1-\tilde{W}(s)]\right).
\end{equation}
This gives an exact expression for the spatial propagator for any distribution of refractory times. One can indeed verify that the propagator is normalized $ \hat{\tilde{\wp}}_r(0,s|v_0) = 1/s$. Furthermore, in the case without any refractory period $W(\tau) = \delta(\tau)$, we recover previous expressions for the propagator under velocity resets.

If one cares only about late-time dynamical behavior, one may naively expect that this reduces to a simple continuous time random walk (CTRW). In this case, the propagator in Fourier-Laplace space is given by the famous Montroll-Weiss formula, which relates the propagator to the distribution of jump lengths and the distribution of the waiting times in between each jump \cite{montroll1965random}. The classical Montroll-Weiss theory assumes independent jump lengths and waiting times, although extension to correlated cases have been considered \cite{liu2013continuous}. In the present case, the waiting time is given by the combination $t_E+\tau$, with $t_E$ the exploration time and $\tau$ the duration of a refractory phase. The jump length however, is determined both by the underlying propagator in the exploration phase and the duration of the exploration phase. Hence the jump lengths depend only on the time $t_E$ of the exploration phase, not the duration of the refractory times. Therefore velocity resetting with refractory times can be seen as a form of CTRW with a particular correlation between the jump lengths and the waiting times. Of course, another strong contrast to the CTRW case is that the dynamics studied here is fully time-resolved and does not perform sudden discrete jumps.

\subsection{Mean-squared displacement}
The moments of position can be obtained from Eq. (\ref{eq:exact}) by differentiation as before, through Eq. (\ref{eq:mom1}). Assuming we are dealing with a process where the first moment always vanishes, which is typically the case if $v_0 =0$, we can proceed as in earlier sections to find
\begin{equation}\label{eq:msd}
     \widetilde{\langle x^2|0,0\rangle}_s  =  \widetilde{\langle x^2 |0,0\rangle}_{s+r}^{(0)} \frac{1 +\frac{r}{s}[1-\tilde{W}(s)]}{\left[ 1 -\frac{r \tilde{W}(s)}{s+r}\right]^2}.
\end{equation}
We consider two scenarios next.

\subsubsection{Mean refractory period is finite $(\langle \tau \rangle <\infty)$:}
As before, the late-time dynamics can be obtained by considering the small-$s$ limit of the Laplace transformed mean squared displacement in Eq. (\ref{eq:msd}). For small $s$, we can approximate $\tilde W(s) = 1-s \langle \tau \rangle + ...$, in which case the dominant singularity of the MSD is $s^{-2}$:
\begin{equation}
     \widetilde{\langle x^2|0,0\rangle}_s  = \frac{r^2  \widetilde{\langle x^2 |0,0\rangle}_{r}^{(0)} }{s^2\left( 1 + r \langle \tau \rangle\right)}.
\end{equation}
Hence the diffusion is normal, with an effective diffusion coefficient
\begin{equation}
    D_\text{eff} = \lim_{t\to \infty} \frac{\langle x^2(t)|0,0 \rangle}{2 t} = \frac{r^2  \widetilde{\langle x^2|0,0 \rangle}_{r}^{(0)} }{ 2 + 2 r \langle \tau \rangle}.
\end{equation}
We see that refractory times suppress the effective diffusion, and when $ \langle \tau \rangle=0$ we recover previous results.

\subsubsection{Mean refractory period is infinite $(\langle \tau \rangle = \infty)$:}
If the mean refractory time diverges due to  power-law distributed refractory periods $W(\tau) \sim 1/(\tau^{1+\alpha})$ with $\tilde W(s) \sim 1- a s^\alpha$, for some constant $a$ and $\alpha \in(0,1)$, we in stead find the small-$s$ behavior
\begin{equation}
     \widetilde{\langle x^2|0,0\rangle}_s  = \frac{r  \widetilde{\langle x^2|0,0 \rangle}_{r}^{(0)} }{ a s^{1+\alpha}}.
\end{equation}
In this case the MSD in real-time reads
\begin{equation}
     \langle x^2(t)\rangle  \simeq \frac{r  \widetilde{\langle x^2|0,0 \rangle}_{r}^{(0)} }{ a } t^\alpha,
\end{equation}
with $\simeq$ denoting equality at late times. Hence the particle will exhibit anomalous diffusion of the sub-diffusive type independently of its underlying dynamics, simply as a consequence of the long refractory periods.

\section{Conclusion and outlook}\label{sec:concl}

 In this paper we have studied stochastic resetting in coupled systems, with particular focus on position and velocity. A general result was derived for the propagator for an observable variable under the indirect effect of resetting of a hidden variable. From this we derived a general recursive equation for the moments, from which one can in principle fully characterize the marginalized process. We applied the proposed framework to transport processes of inertial particles where velocity undergoes resetting, and show that generically the late-time dynamics shows normal diffusion even if the reset-free system is anomalous. This we attribute to the tempering effect stochastic resetting has on velocity auto-correlation functions. We derived a compact expression for the effective drift and diffusivity coefficients. When a temporal crossover between two dynamical regimes exists in the underlying dynamics, this translates into a crossover as a function of resetting rate in the effective diffusivity, where different scaling behaviors are observed. We tested the validity of our predictions in the cases of underdamped Brownian motion and for the random acceleration process, both showing excellent agreement with simulated data. Extension to the case where refractory periods are included after each reset was also considered, in which case resetting-induced anomalous diffusion can be observed. 

The main results of this paper are based on two crucial assumptions. First is the homogeneity in the variable $x$ that is not reset, and second is the full renewal structure represented by Eq. (\ref{eq:firstrenew}). Extensions of the results presented here to account for either spatial heterogeneity or for example non-renewal reset structure such as in \cite{bodrova2019nonrenewal} would be very interesting, shining light on anomalous diffusion processes with velocity resets in systems with quenched or annealed disorder. Generalizations to more complex types of resetting, such as proportional resetting \cite{pierce2022advection} or non-Poissonian waiting times could also be considered.
 
 Finally, it would also be interesting to investigate resetting in coupled systems from a thermodynamic perspective. Recently much effort has been put into understanding the stochastic thermodynamics of resetting, with both entropy production and work having been considered \cite{gupta2022work, mori2023entropy,olsen2023thermodynamic}. However, in the present model the observable variable is not the one undergoing resets. Hence it would be interesting to investigate bounds on the thermodynamic cost based on partial accessible information, a topic which has gained considerable attention in the past decade \cite{roldan2010estimating, amann2010communications,polettini2017effective,bilotto2021excess, ehrich2021tightest,neri2022estimating,ghosal2022inferring,pietzonka2023thermodynamic, baiesi2023effective,ghosal2023entropy}.

\section*{Acknowledgments}
Insightful discussions and interactions with Kevin Pierce and Deepak Gupta are gratefully acknowledged. The authors acknowledge support from the Deutsche Forschungsgemeinschaft (DFG) within the project LO 418/29-1.


\providecommand{\newblock}{}

\end{document}